\documentclass[letterpaper,12pt]{extarticle}  

\usepackage{import}
\usepackage{local}
\usepackage[left=.75in,top=.75in,bottom=.75in,right=.75in]{geometry}

\usepackage[utf8]{inputenc}
\usepackage[T1]{fontenc}

\usepackage{float}
\usepackage{physics}
\usepackage{graphicx}
\usepackage{dcolumn}
\usepackage{bm}
\usepackage{mathrsfs}
\usepackage{amssymb}
\usepackage{amsmath}
\usepackage[normalem]{ulem}
\usepackage{wrapfig}
\usepackage[compress,square,sort,comma,numbers]{natbib}
\usepackage{mathtools}

\usepackage{hyperref}
\usepackage{bookmark}
\usepackage{lipsum}

\hypersetup{%
  breaklinks=true,
  colorlinks=true,
  linkcolor=black,
  filecolor=black,
  urlcolor=black,
  citecolor=black,
  pdfborder=0 0 0,
}

\begin{document}

\title{\LARGE Motor crosslinking augments elasticity in active nematics}
\author{%
\textbf{Steven A. Redford\textcolor{Accent}{\textsuperscript{1,2}}, %
Jonathan Colen\textcolor{Accent}{\textsuperscript{3}}, %
Jordan L. Shivers\textcolor{Accent}{\textsuperscript{4,5}}, %
Sasha Zemsky\textcolor{Accent}{\textsuperscript{6}}, %
Mehdi Molaei\textcolor{Accent}{\textsuperscript{7}}, %
Carlos Floyd\textcolor{Accent}{\textsuperscript{4}}, %
Paul V. Ruijgrok\textcolor{Accent}{\textsuperscript{6}}, %
Vincenzo Vitelli\textcolor{Accent}{\textsuperscript{3,4}}, %
Zev Bryant\textcolor{Accent}{\textsuperscript{6,8}}, %
Aaron R. Dinner\textcolor{Accent}{\textsuperscript{2,4,5,*}}, %
Margaret L. Gardel\textcolor{Accent}{\textsuperscript{2,3,4,7,*}}%
}\\
\begin{small}
\textcolor{Accent}{\textsuperscript{1}}The Graduate Program in Biophysical Sciences, University of Chicago, Chicago, IL 60637, U.S.A. \\ 
\textcolor{Accent}{\textsuperscript{2}}Institute for Biophysical Dynamics, University of Chicago, Chicago, IL 60637, U.S.A. \\ 
\textcolor{Accent}{\textsuperscript{3}}Department of Physics, University of Chicago, Chicago, IL 60637, U.S.A. \\
\textcolor{Accent}{\textsuperscript{4}}James Franck Institute, University of Chicago, Chicago, IL 60637, U.S.A. \\ 
\textcolor{Accent}{\textsuperscript{5}}Department of Chemistry, University of Chicago, Chicago, IL 60637, U.S.A. \\
\textcolor{Accent}{\textsuperscript{6}}Department of Bioengineering, Stanford University, Stanford, CA, 94305, U.S.A. \\ 
\textcolor{Accent}{\textsuperscript{7}}Pritzker School of Molecular Engineering, The University of Chicago, Chicago, IL, 60637, U.S.A. \\ 
\textcolor{Accent}{\textsuperscript{8}}Department of Structural Biology, Stanford University School of Medicine, Stanford, CA, 94305, U.S.A. \\
\textcolor{Accent}{\textsuperscript{*}}Correspondence: \textcolor{Accent}{dinner@uchicago.edu, gardel@uchicago.edu} \\ \end{small}
}

\maketitle

\begin{onehalfspacing}

\section{abstract}

In active materials, uncoordinated internal stresses lead to emergent long-range flows. 
An understanding of how the behavior of active materials depends on mesoscopic (hydrodynamic) parameters is developing, but there remains a gap in knowledge concerning how hydrodynamic parameters depend on the properties of microscopic elements.
In this work, we combine experiments and multiscale modeling to relate the structure and dynamics of active nematics composed of biopolymer filaments and molecular motors to their microscopic properties, in particular motor processivity, speed, and valency. We show that crosslinking of filaments by both motors and passive crosslinkers not only augments the contributions to nematic elasticity from excluded volume effects but dominates them. By altering motor kinetics we show that a competition between motor speed and crosslinking results in a nonmonotonic dependence of nematic flow on motor speed. By modulating passive filament crosslinking we show that energy transfer into nematic flow is in large part dictated by crosslinking. Thus motor proteins both generate activity and contribute to nematic elasticity. Our results provide new insights for rationally engineering active materials.

\newpage
\section{Introduction}

Systems composed of active agents that locally break detailed balance can exhibit striking collective behaviors that are inaccessible to assemblies that couple to energy sources in a nondirected fashion (e.g., thermally)
\cite{ramaswamy_mechanics_2010,marchetti_hydrodynamics_2013}. 
These behaviors include directed collective motion, enhanced information storage, giant number fluctuations, self-sorting, and motility-induced phase separation \cite{sanchez_spontaneous_2012,del_junco_energy_2018,kumar_trapping_2019,ramaswamy_active_2003}.
A better understanding of active systems can suggest mechanisms in natural systems \cite{cavagna_bird_2014,dombrowski_self-concentration_2004,wensink_meso-scale_2012}, enable control of nonequilibrium pattern formation, and guide the design of new materials 
\cite{guillamat_control_2016,wu_transition_2017,ross_controlling_2019,zhang_spatiotemporal_2021}.
One of the most well-studied classes of active materials is active nematics (also known as active  liquid crystals)
\cite{kumar_catapulting_2022,zhang_spatiotemporal_2021}. In nematics, elongated components (mesogens) interact locally through excluded volume yielding a material that exhibits long-ranged orientational order while maintaining translational fluidity \cite{gennes_physics_1993,doostmohammadi_onset_2017}. The tendency of the mesogens to align gives a nematic an effective elasticity that resists distortions and acts to align the field as a whole \cite{marchetti_hydrodynamics_2013}. This tendency to align is opposed by activity (i.e., mechanical work done on the individual elements of a system). Activity induces structural distortions and flow in the nematic field leading to a dynamical steady state.

There are various ways to characterize structure in an active nematic. These include the spacing of topological defects, the correlation length of the orientation of the mesogens (director field), and the correlation length of the velocity or vorticity \cite{giomi_geometry_2015,hemingway_correlation_2016,doostmohammadi_active_2018}.  However, theory \cite{giomi_geometry_2015}, simulation \cite{hemingway_correlation_2016}, and experiments \cite{kumar_tunable_2018,lemma_statistical_2019} suggest that these quantities all scale identically with activity---i.e., for a given set of conditions, active nematic dynamics are governed by a single length scale, $\ell$.  This length scale arises from the balance of the elastic stress, $K/\ell^2$, where $K$ is the elastic constant, with the active stress scale, $\alpha$, so that $\ell=\sqrt{K/\alpha}$ \cite{hemingway_correlation_2016}. While $\ell$ quantifies how much energy imparted by activity is stored in distortions to the nematic field, the average flow speed of the nematic captures how much energy is dissipated viscously. As such, by force balance, the average flow speed in a nematic is expected to scale as \(v \sim \alpha\ell/\eta\sim \sqrt{K \alpha}/\eta\), where $\eta$ is the solvent viscosity \cite{hemingway_correlation_2016}. Thus exerting control over $K$ and $\alpha$ affords control over the steady-state dynamics and structure of an active nematic.

How exactly $K$ and $\alpha$ relate to microscopic properties of the elements that make up active nematics is not well understood.  In active nematics composed of cytoskeletal elements---semiflexible filaments, molecular motors, and crosslinkers---activity is generated when the molecular motors hydrolyze adenosine triphosphate (ATP) and slide pairs of filaments, giving rise to interfilament strain and extensile force dipoles \cite{sanchez_spontaneous_2012}. Biochemical regulation affords control of microscale mesogen properties and active stresses allowing for explicit tuning of hydrodynamic properties on a microscopic scale.
For example, in active nematics composed of actin filaments and myosin II motors, the elastic constant was shown to depend on filament length \cite{kumar_tunable_2018,kumar_catapulting_2022}. 
In nematics composed of microtubules and kinesin, active stresses have been modulated by changing the concentration of ATP ([ATP]) available to motors.   
In this case, the impact of altering [ATP] was to affect the activity through motor stepping speed and not the elasticity \cite{lemma_statistical_2019}. The motor employed in this and other studies of cytoskeletal active nematics (kinesin and myosin II filaments) have high processivities. That is, they almost never detach from filaments before reaching their ends \cite{verbrugge_novel_2009,bloemink_shaking_2011}. 
Because a motor must link a pair of filaments to generate extensile stress, one would expect that differences in filament binding propensities lead to differences in force transmission capabilities. 
Indeed, filament crosslinking was observed to impact local rigidity and force transmission in other cytoskeletal contexts \cite{stam_filament_2017}.  However, the roles of motor processivity and, more generally, crosslinking in active nematics have not been explored to the best of our knowledge.

To address this gap, here we utilize synthetic myosin motors that range in their propensities for binding filaments \cite{schindler_engineering_2014}. We tune processivity through both [ATP] and motor oligomerization state (valency). We find that nematic speed depends nonmonotonically on [ATP], reflecting opposite trends in filament strain and crosslinking with [ATP].  We find that crosslinking modulates the elasticity, and we introduce a simple model that accounts for the observed trends. Consistent with the model, we show that the addition of the passive crosslinker filamin also modulates elasticity and in so doing alters the energetic balance in active flows. Our results reveal a previously unappreciated connection between activity and elasticity through motor proteins and show how these quantities can be tuned independently through molecular composition.

\section{Results}

To probe how the microscopic interactions between a motor and filament control nematic structure and dynamics, we pair \textit{in vitro} experiments with multiscale modeling. Experimentally, we can alter processivity by changing the availability of ATP or motor valency. Specifically we employ synthetic myosin motors that consist of the enzymatic head from {\it Chara} myosin XI, which is linked via a flexible linker to an engineered multimerization domain \cite{schindler_engineering_2014}. By utilizing different multimerization domains, either engineered GCN4 coiled-coils \cite{harbury_switch_1993} or de novo two-helix hairpins \cite{xu_computational_2020}, which form clusters of well-defined sizes, we are able to query clusters with identical enzymology but with three, four, or eight heads (Fig.\ 1A). In the high ATP limit the {\it Chara} myosin XI head has a low duty ratio, meaning it spends less than half of its catalytic cycle bound to an actin filament \cite{ito_kinetic_2007,sumiyoshi_insight_2007,haraguchi_discovery_2022}. Because this duty ratio depends on [ATP] motor velocity and the distance a motor travels before dissociating (run length) on single filaments also depend strongly on [ATP]:  at \(\text{[ATP]}=10\ \mu\)M, tetrameric clusters have single-filament velocities of \(0.5\ \mu \text{m\ s}^{-1}\) with run lengths of \(4\ \mu\)m, while at \(\text{[ATP]}=500\ \mu\)M, the velocity is \(10\ \mu \text{m\ s}^{-1}\), and the run length is \(0.5\ \mu\)m (Fig.\ S1) \cite{schindler_engineering_2014}. 

\subsection{A microscopic model relates motor properties to hydrodynamic parameters.}

To understand how the activity depends on [ATP] in our system and in turn to make predictions for the nematic speed and correlation length through the relations $v\sim\sqrt{K\alpha}$ and $\ell\sim \sqrt{K/\alpha}$, we developed a microscopic model of motors with variable valencies.  
Because activity is generated via filament pair strain rate and not merely motor speed, this model focuses on the calculation of filament strain rate, $\varepsilon$. We then use this quantity in the scaling relation $\alpha\sim \varepsilon^\beta$, which was previously observed to hold for active nematics composed of microtubules and kinesin motors \cite{lemma_statistical_2019}, given the known dependence on [ATP] of the speed of single kinesin motors walking on single filaments \cite{verbrugge_novel_2009}.

Building upon a previous approach \cite{vilfan_elastic_2005}, we coarsely approximate the catalytic cycle of each head using three states:  (1) unbound from the filament with ATP, (2) bound to the filament in the post-powerstroke state with ADP, 
and (3) bound to the filament without a nucleotide (Fig.\ 1B).  Transitions between these states are irreversible. An essential idea is that a head with ATP has low affinity for the filament.  As a result, the transition from state 1 to state 2 requires ATP hydrolysis.  Similarly, the head quickly releases the filament once it exchanges ADP for ATP, and the rate of the transition from state 3 to state 1 is linearly dependent on [ATP]. We simulate the cycle for each head independently. That is, if there are $n$ heads in a simulation, there are $3n$ states to keep track of. Because the heads are independent and rates are irreversible, there are only $n$ allowed transitions at any time. To evolve the system forward, we perform the Gillespie algorithm over all possible transitions at a given time \cite{gillespie_exact_1977}. This scheme allows us to simulate clusters of independent heads with any valency.

We assume that the joint between the lever arm and the multimerization domain is flexible and that the motor prefers to bind in its least strained position. Thus, when a head undergoes a transition from state 1 to state 2 and binds to a filament, we draw its position from the normal distribution \(N(x(t)+s/2,s/2)\). Here, \(x(t)\) is the position of the multimerization domain that couples independent heads together and $s$ is the average step length of a motor. On each filament, we take $x(t)$ to be a distance $s/2$ ahead of the rearmost bound head. Assuming fast diffusion relative to binding rates, when a motor can bind multiple filaments we choose between them randomly with equal probability. When a transition occurs, $x(t)$ is reevaluated. We calculate the average velocity of a motor on a filament as the total distance a motor travels divided by the final time in the simulation. For pairs of filaments, strain is only recorded if motion occurs while the motor crosslinks the two filaments \cite{marchetti_hydrodynamics_2013,gao2015multiscale}. We compute the filament strain rate, $\varepsilon$, by dividing the total strain by the final time in the simulation. We also compute the probability of crosslinking, $P_\text{cl}$, as the fraction of time that both filaments are bound simultaneously.

We scan the three rate constants ($k_{12}$, $k_{23}$, $k_{31}$) to identify values that yield average single-filament speeds and run lengths (i.e., the length traveled between the first time a head is bound to the last time) that reproduce measured trends and approximately correspond to measured values from experiments with tetrameric clusters (Fig.\ S1) \cite{schindler_engineering_2014}. Two filament results, $\varepsilon$ and $P_\text{cl}$, for a tetrameric motor cluster are shown in Figs.\ 1C,D. These simulations show that \(P_\text{cl}\), decreases while  \(\varepsilon\) increases with [ATP].

As described above, we use the computed strain rate to estimate the activity by $\alpha\sim \varepsilon^\beta$.  We use \(\beta = 0.1\) to account for the flexibility of the synthetic myosin XI motor \cite{schindler_engineering_2014} (for comparison, values ranging from 0.31 to 1.54 are considered for kinesin in \cite{lemma_statistical_2019}).  Substituting the resulting $\alpha$ into $v\sim\sqrt{K\alpha}$ and $\ell\sim\sqrt{K/\alpha}$, we obtain an increase in \(v\) and a decrease in \(\ell\) with [ATP], for fixed $K$ (Fig.\ 1E). 

\begin{figure}[bt!]
    \centering
    \includegraphics[width=0.75\textwidth]{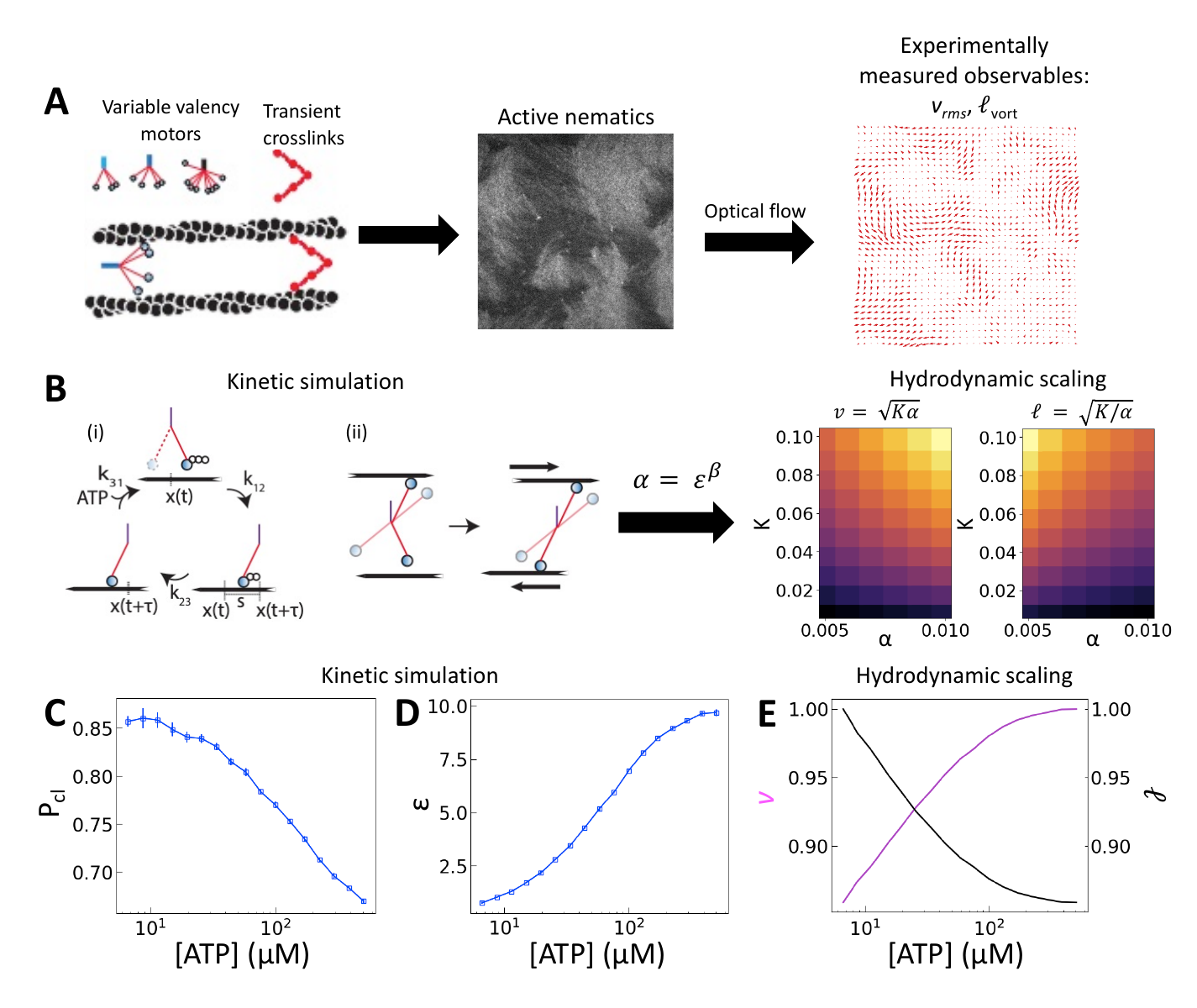}
    \caption{\textbf{[ATP] and activity can be related through a microscopic model.} 
    (A) Schematic of the experiments. We study synthetic motors with controlled numbers of myosin XI enzymatic heads that bind and slide actin filaments of length 2 $\mu$m at an oil-water interface. Due to the polarized binding of a dye to actin filaments, regions with filaments oriented vertically in the laboratory frame appear brighter than those oriented horizontally \cite{kumar_catapulting_2022,kumar_tunable_2018}. The experimental images are analyzed by optical flow \cite{sun_secrets_2010} to estimate the horizontal and vertical components of the velocity at each pixel.  From the velocity field, we calculate the average flow speed, \(v_\text{rms}\), and average vortex radius \(\ell_\text{vort}\) as in \cite{molaei_measuring_2023}. (B) We simulate the catalytic cycle of myosin XI with three states: (1) unbound with ATP (top), (2) bound with ADP (right), and (3) bound while nucleotide free (left). (i) Rate constants are tuned based on prior measurements of speed and processivity on single filaments (Fig.\ S1). (ii) We extend the simulation to two filaments as described in the text and compute the filament extension rate, \(\varepsilon\), and the  probability of crosslinking, \(P_\text{cl}\), as described in the text. These quantities are used to compute the nematic speed and correlation length as $v=\sqrt{K\alpha}$ and $\ell=\sqrt{K/\alpha}$, respectively. (C) \(P_\text{cl}\) and (D) \(\varepsilon\) from two-filament simulations for a cluster with four heads. (E) Normalized \(v\) (magenta) and \(\ell\) (black) for activity derived from (D) assuming constant elasticity, \(K = 0.001\).
    }
    \label{figure1}
\end{figure}

\subsection{Nematic elasticity depends on the probability of crosslinking.}

To test our predictions, we use nematics composed of short (2 $\mu$m) actin filaments labelled with tetramethylrhodamine (TMR) and synthetic motors with {\it Chara} myosin XI enzymatic heads \cite{schindler_engineering_2014}.  We form nematics by crowding the actin filaments to a surfactant stabilized oil-water interface through depletion forces imposed by methyl-cellulose (Fig.\ 1A). Once the nematic is formed, we add \(120\) pM tetrameric motors to the sample to introduce activity. We image the sample with time-lapse fluorescence microscopy at a rate of 0.5 frames/s for 100 s. Because of the polarization of TMR dye along filaments and the polarization of our excitation laser, brighter (darker) patches represent filaments oriented vertically (horizontally) in the imaging plane \cite{kumar_catapulting_2022,kumar_tunable_2018}.
Given the video microscopy data, we estimate the nematic velocity at each pixel using optical flow \cite{sun_secrets_2010}, as described in Materials and Methods. 

The results for one series of [ATP] are shown in Fig.\ 2.  As we expected, the length scale \(\ell_\text{vort}\),  calculated using correlated displacement velocimetry, decreases as [ATP] increases (Fig.\ 2A,C) \cite{molaei_interfacial_2021}. We use \(\ell_\text{vort}\) to quantify length scale because it agrees well with the velocity correlation length but requires fewer assumptions to measure \cite{molaei_measuring_2023,hemingway_correlation_2016} (Fig.\ S2). While \(\ell_\text{vort}\) decreases with [ATP], the root mean square flow velocity, \(v_\text{rms}\), exhibits a nonmonotonic dependence on [ATP], with a peak at \(50\ \mu\)M (Fig.\ 2A,B). This behavior contrasts with the model prediction (Fig.\ 1E), suggesting that something is missing from the model.  

Given previous work in which material elasticity depends on the concentration of crosslinkers \cite{ahmadi_hydrodynamics_2006,kruse_asters_2004}, we reasoned that the elastic constant $K$ should depend (linearly) on the effective concentration of crosslinkers, $c_e$:
\begin{equation}\label{eqn:K}
    K \sim K_0 + \kappa c_e,
\end{equation}
where \(K_0\) is the baseline nematic elastic modulus that arises from excluded volume interactions between filaments \cite{zhang_interplay_2017,ahmadi_hydrodynamics_2006}, and \(\kappa\) represents the energetic penalty for filament deformation at a given concentration of crosslinker. 
Here, because the only crosslinkers are motors, we expect \(c_e = c_m P_\text{cl}\), where $c_m$ is the concentration of motors which is taken to be 1 throughout this work.  Using \eqref{eqn:K} for $K$ with $P_\text{cl}$ from the simulation in the scaling relations $v\sim\sqrt{K\alpha}$ and $\ell\sim\sqrt{K/\alpha}$, we obtain nonmonotonic $v$ and decreasing $\ell$ with increasing [ATP] (Fig.\ 2D,E).  Physically, there is a competition between the tendency for increased [ATP] to increase motor speed, resulting in a higher strain rate, and to reduce motor binding, resulting in lower $P_\text{cl}$.  In the case of kinesin, the latter tendency is negligible due to biochemical coupling and thus was not necessary to consider in previous studies \cite{lemma_statistical_2019,verbrugge_novel_2009}.

The peak in $v$ becomes more pronounced as the second term in \eqref{eqn:K} becomes large compared with the first (Fig.\ 2D). 
To understand how a peak in $v$ could arise from these scaling relationships, we differentiate \(v=\sqrt{K\alpha}\) with respect to [ATP] and solve for the maximum by setting the resulting expression equal to zero.  This yields
\begin{equation}\label{eqn:vmax}
    \alpha_\text{peak} = -\frac{K\alpha'}{K'},
\end{equation}
where $\alpha_\text{peak}$ is the activity that corresponds to the maximum velocity, and $K'$ denotes a derivative with respect to [ATP]. Note that because $P_\text{cl}$ always decreases with [ATP], $K' \le 0$.
For a fixed dependence of the strain rate and thus the activity on [ATP], larger $\kappa$ results in larger $K'$ relative to $K$ and thus smaller $\alpha_\text{peak}$ (i.e., $\alpha_\text{peak}$ at lower [ATP]).  Consistent with this reasoning, the peak in Fig.\ 2D moves to the left as $\kappa$ increases.  It is also worth noting here that changes in \(\beta\) affect the balance in this equation as well. If we increase \(\beta\), the nematic speed increases monotonically with [ATP], similar to a decrease in \(\kappa\) (Fig.\ S3). As such we set \(\kappa = 10 K_0\) and \(\beta = 0.1\) for the rest of this work.

Note that the variations in \(v\) and \(\ell\) are smaller than in experiment. This is a reflection of our simplifying assumptions in this model. On a hydrodynamic scale, we assume that turbulent scaling relations hold at all concentrations, even though we expect them to only hold above a critical [ATP]. Furthermore, our assumption that $K$ is linear in \(P_\text{cl}\) is likely an oversimplification. Microscopically we neglect complex coupling \cite{forgacs_kinetics_2008} and correlated binding \cite{vilfan_elastic_2005} in our motor stepping model, both of which would reduce \(P_\text{cl}\) at high [ATP]. The model can readily be tuned to adjust for these assumptions, but we do not pursue that here for simplicity.

\begin{figure}[bt!]
    \centering
    \includegraphics[width=0.75\textwidth]{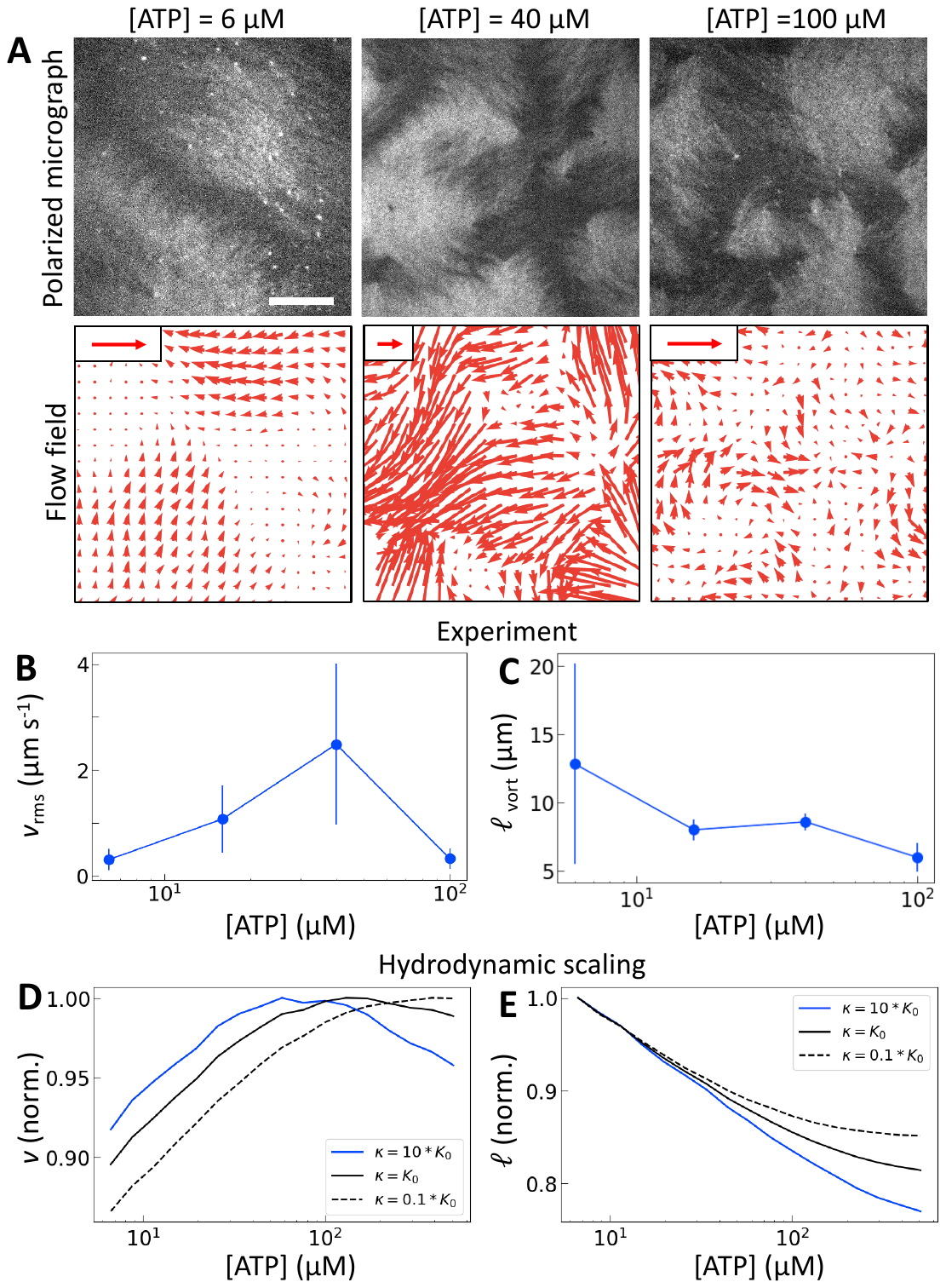}
    \caption{\textbf{Motor crosslinking modulates nematic elasticity.} 
    (A, top row) Polarized fluorescence micrographs of nematics (gray scale) driven by tetrameric motor clusters from \cite{schindler_engineering_2014} with [ATP] of 6, 40 or 100 \(\mu \text{M}\) (concentration of motors is 120 pM). (A, bottom row) Velocity fields estimated from optical flow. Scale arrows are \(3\  \mu \text{m/s}\). (B) Average flow speed, \(v_\text{rms}\), for the experiments in (A) and similar ones with [ATP] of 16 $\mu$M. Error bars are standard deviations of speed over 100 s of steady-state activity. (C) Critical vorticity length scale, \(\ell_\text{vort}\), measured as in \cite{molaei_measuring_2023}, for the same experiments as in (B). Error bars are standard deviations on 5 sets of 5 non-overlapping frames. (D and E) Normalized \(v\) and $\ell$ for tetrameric motors calculated from the model scaling with various ratios of \(\kappa\) to \(K_0\). All calculations presented subsequently use \(\kappa = 10 K_0\) and \(\beta = 0.1\).
    }
    \label{figure2}
\end{figure}

\subsection{Motor valency tunes nematic dynamics.}

We now consider how the motor valency (i.e., the number of heads in a cluster) affects the structure and dynamics of the active nematics. Simulations of motors on single filaments show that increasing the motor valency reduces the speed and increases the processivity, consistent with experimental measurements \cite{schindler_engineering_2014} (Fig.\ S4). These trends shift the dependence of \(\varepsilon\) on [ATP]  in simulations of motors on two filaments such that higher [ATP] is required to reach the same relative extension rate (Fig.\ 3A). Higher valency also leads to a greater probability of crosslinking across all ATP concentrations and a smaller relative decrease in crosslinking across the range of [ATP] that we consider (Fig.\ 3B). 
These microscopic trends lead to a valency-dependent shift of the peak in \(v\) to higher [ATP] (Fig.\ 3C, dotted line) 
and a decrease in the relative change in \(\ell\) between low and high [ATP] (Fig.\ 3D).

Experimentally, we utilize the control afforded by the motor's multimerization domain to consider clusters with \(n = 3, 4\), or \(8\) heads. We take into account the contributions of cluster valency and total number of motor heads by considering trimeric and tetrameric motor clusters at \(120\) pM and octameric motor clusters at \(60\) pM (Fig.\ 3E). This allows us to separate the contributions from cluster valency and the total head number in the system. We find that the peak in \(v_\text{rms}\) is indeed dependent on cluster valency and shifts to higher [ATP] as valency increases (Fig.\ 3E). This trend holds across multiple independent series (Fig.\ S5). In fact, the shift that we find in experiment closely matches that predicted by our simulations (Fig.\ 3C). Furthermore, as valency increases, \(\ell_\text{vort}\) at a given [ATP] increases (Fig.\ 3F). Thus we can access different ATP response regimes in these nematics by tuning motor valency. However, separating the contributions of \(P_\text{cl}\) and \(\varepsilon\) in these experiments is not possible as these quantities vary simultaneously as valency changes (Fig.\ 3A,B). 

\begin{figure}[bt!]
    \centering
    \includegraphics[width=0.75\textwidth]{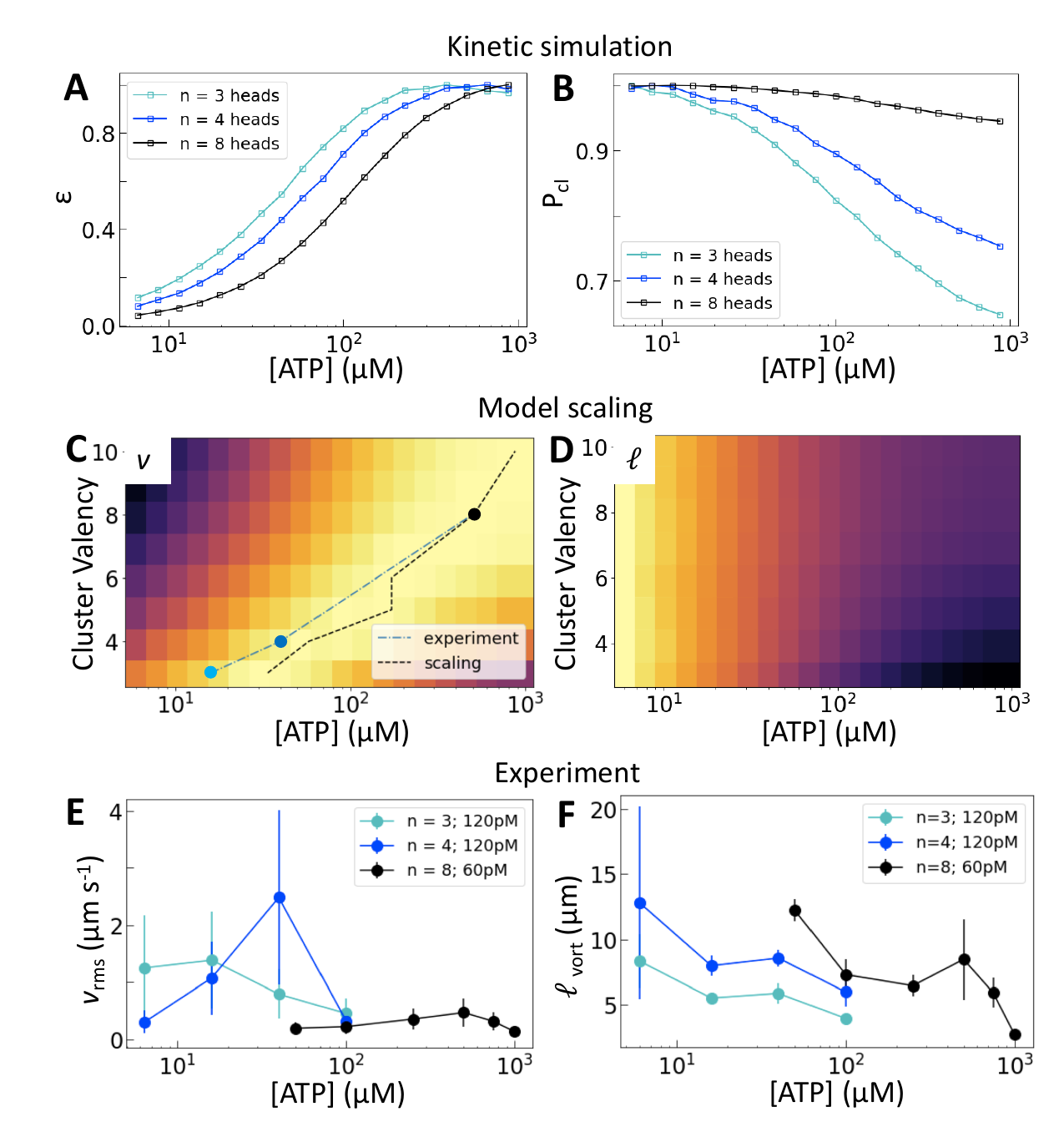}
    \caption{\textbf{Motor valency tunes nematic dynamics.} 
     (A and B) Normalized \(\varepsilon\) and \(P_\text{cl}\) calculated for clusters of variable valency. (C and D) Normalized \(v\) and \(\ell\) from model scaling. The black dotted line in (C) traces the location of the peak in nematic speed; symbols show the positions of peak speeds in (E). Brighter colors are higher values. (E and F) \(v_\text{rms}\) and \(\ell_\text{vort}\) for a range of ATP concentrations and cluster valencies. Error bars are standard deviations on 5 sets of 5 non-overlapping frames from a single experiment. Independent replicates are found in Fig.\ S5.
    }
    \label{figure3}
\end{figure}

\subsection{Crosslinking modulates the efficiency of nematic energy transfer.}%

To separate the effects of crosslinking and strain rate, we consider the effects of adding the passive crosslinker filamin (FLN).  Here, we use active nematics driven by trimeric motors because they have the lowest baseline level of crosslinking. 
To incorporate the contribution from passive crosslinkers in the model, we simply add a contribution to the effective concentration of crosslinkers: \(c_e = c_m P_{cl} + c_p\),  where \(c_p\) is the concentration of passive crosslinkers.  Otherwise the model is the same (Fig.\ 4A). 
This model predicts that the addition of passive crosslinkers leads to a shift in the peak in $v$ to higher [ATP] (Fig.\ 4B). We note that this shift is different from that in response to changing the valency in that it occurs for constant \(\varepsilon\) and \(P_{cl}\). Experimentally, we find that adding crosslinker to these samples yields a dramatically longer length scale as is expected from increased \(K\) (Fig.\ 4C,E). Furthermore, we find that increased concentrations of passive crosslinker do indeed lead to a shift in the peak in \(v_\text{rms}\) to higher [ATP] (Fig.\ 4D,E). These observations support our model, in which crosslinking linearly increases the elastic modulus.
In turn, the shift in \(v_\text{rms}\) can be understood in terms of \eqref{eqn:vmax}. Previously we discussed the case of increasing \(\kappa\), which increases \(K'\), shifting $\alpha_\text{peak}$ to lower [ATP]. By contrast, adding passive crosslinkers leaves \(K'\) unchanged while increasing overall $K$, shifting $\alpha_\text{peak}$ to higher [ATP]. 

As noted before this shift is accompanied by an increase in \(\ell_\text{vort}\) and \(v_\text{rms}\) (Fig.\ 4C,D). Thus for a given [ATP] the nematic features fewer defects but moves faster (Fig.\ 4E). These changes occur without a substantial change in \(\varepsilon\), suggesting that shifts in \(K\) affect how the activity supplied by motors manifests in nematic dynamics. Indeed, lattice Boltzmann simulations show that in the high activity regime total energy in the nematic actually increases with \(K\) (Fig.\ S6). This indicates a crucial role for filament crosslinking in determining the efficiency of energy transfer from motor stress into active nematic motion.

\begin{figure}[bt!]
    \centering
    \includegraphics[width=0.75\textwidth]{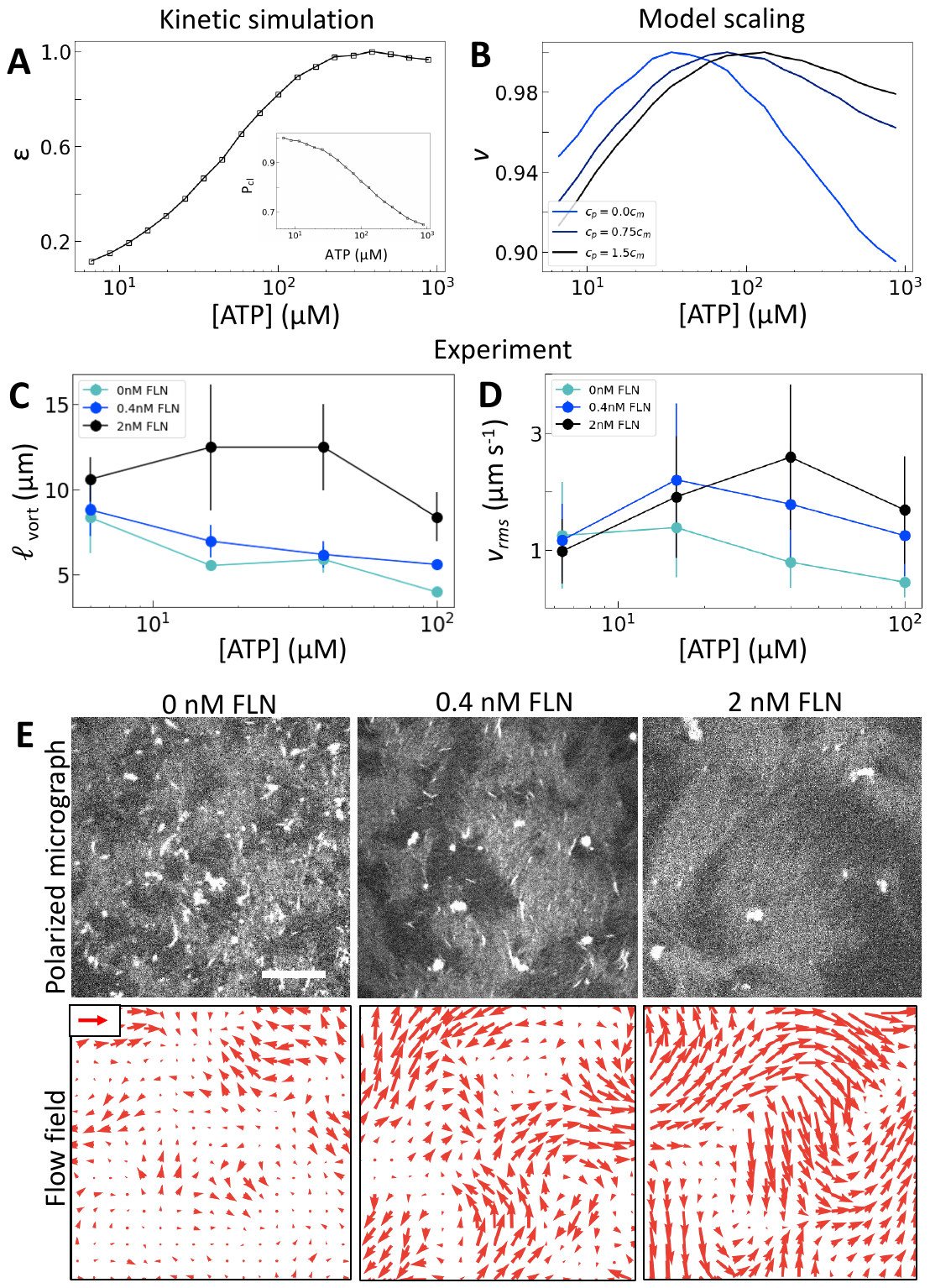}
    \caption{\textbf{Microscopic crosslinking alters nematic energy distribution.} 
    (A) Normalized \(\varepsilon\) and \(P_\text{cl}\) (inset) calculated for trimeric motors. (B) Normalized \(v\) from model scaling. (C and D)  \(v_\text{rms}\) and \(\ell_\text{vort}\) measured for trimeric driven nematics with filamin (FLN) added as indicated. (E) Polarized fluorescence micrographs (gray, top row) with corresponding flow fields (red arrows, bottom row) for trimeric motors at 100 \(\mu \text{M}\) ATP with FLN added as indicated. Scale arrow is 3\(\mu \text{m/s}\).
    }
    \label{figure4}
\end{figure}

\section{Conclusions}

In this work we showed that crosslinking has a profound effect on active nematic dynamics through elasticity. Previous work with high processivity motors focused on the motors' role in activity despite clues to their role in elasticity from machine learning \cite{colen_machine_2021} and experiments in the low [ATP] limit  \cite{lemma_statistical_2019}. Our investigation here of active nematics with low processivity motors revealed that reduced filament crosslinking at high [ATP] leads to reduced nematic elasticity and a nonmonotonic dependence of nematic speed on [ATP]. 
Indeed, we find that the contribution to elasticity from crosslinking dominates that from excluded volume interactions. We expect this to be the case even in cytoskeletal active nematics in which crosslinking is constant across [ATP], as in active nematics composed of microtubules and kinesin motors \cite{verbrugge_novel_2009,lemma_statistical_2019}. 

Our results suggest that exquisite control over active nematics dynamics can be achieved through the choice of molecular composition. Increasing motor valency affects both the activity and the elasticity due to the effects on both the strain rate and filament crosslinking. Adding passive crosslinkers in principle allows one to tune just the elasticity.  
That both motors and crosslinkers affect elasticity has long been appreciated for actin gels \cite{stam_isoforms_2015,gardel_elastic_2004}. 
Transient crosslinkers have also been shown to tune viscoelastic properties in fluid actin droplets \cite{scheff_tuning_2020,weirich_liquid_2017}. 
Our results suggest that the degree that motor proteins dictate elasticity can be tuned by their physical and biochemical properties.
It is thus interesting to speculate that the fantastic diversity of naturally occurring motors and crosslinkers reflects in part evolutionary pressures to achieve different materials properties.

Our study is a step toward quantitatively linking hydrodynamic parameters of active materials to microscopic properties.  How transferable such relations may be is an open question.  For example, even though active nematics composed of  bacteria can be described in the hydrodynamic limit with similar scaling laws, activity is generated by microscopic mechanisms that are distinct from the active nematics considered here \cite{wensink_meso-scale_2012}. As a result, the characters of their force dipoles may also be distinct, despite both being extensile.  While this suggests that it will be necessary to go beyond scaling relations to characterize active materials fully, it is also an opportunity for tailoring active materials with unique properties.

\newpage
\section{Acknowledgements}
We thank Chunfu Xu for sharing the sequence of the octameric helical bundle construct before publication. SAR is grateful to Cristian Suarez and Rachel Kadzik for help purifying proteins and valuable discussions. This work was partially supported by the University of Chicago Materials Research Science and Engineering Center, which is funded by National Science Foundation under award number DMR-2011854. MLG and ZB acknowledge support from NSF award DMR-2215605 and NIH R01GM143792. ARD acknowledges support from NSF Award MCB-2201235. ZB acknowledges support from NIH R01GM114627. SAR was supported by the NIH under award T32 EB009412. Simulations were performed on computational resources provided by the University of Chicago Research Computing Center.

\section{Author Contributions}
SAR, MM, ZB, ARD, and MLG designed the research. SZ and PVR designed motor constructs and expressed motor proteins. SAR, ZB, and ARD designed the kinetic simulation. SAR performed the experiments and kinetic simulations. JC and JLS developed the hydrodynamic connection between simulations and scaling laws. CSF performed lattice Boltzmann simulations. All authors contributed to and approved the manuscript.

\section{Conflicts of Interest}
The authors declare no conflicts of interest.

\section{Materials and Methods}
\subsection{Experimental Procedures}
\subsubsection{Protein Purification}
Monomeric actin was purified from rabbit skeletal muscle acetone powder (Sigma-Aldrich, St. Louis, MO) as described previously \cite{spudich_regulation_1971} and stored in G-buffer [2mM Tris pH 8, 0.2mM ATP, 0.5mM DTT, 0.1mM CaCl2, 1mM NaN3, pH to 8]. Actin was labelled with Tetramethylrhodamine-6-maleamide (TMR; Life Technologies, Carlsbad, CA). F-Actin Capping Protein was purified as described previously \cite{burke_bacterial_2017} and stored in CP buffer [10mM Tris pH 7.5, 40mM KCl, 0.5mM DTT, 0.01\% NaN$_3$, 50\% Glycerol].

\subsubsection{Cloning and purification of motor constructs}
The tetrameric motor construct CM11CD7462R\(\sim\)1R\(\sim\)TET is described in \cite{schindler_engineering_2014}. Motor constructs were assembled from gene fragments encoding the \emph{Chara corallina} myosin XI motor domain (residues 1–746), \emph{Dictyostelium} $\alpha$-actinin (residues 266–502 for the lever arm and residues 266–388 for the flexible linker), a multimerization domain, and a C-terminal HaloTag and Flag Tag (DYKDDDDK). The tetrameric motor construct contains the GCN4 leucine zipper variant p-LI as the multimerization domain, which forms a parallel tetrameric coiled-coil \cite{harbury_switch_1993}. In the trimeric construct, the multimerization domain was replaced with the GCN4 variant p-II, which forms a coiled-coil trimer rather than a tetramer \cite{harbury_switch_1993}, as previously described for similar constructs \cite{schindler_engineering_2014}. To create the octameric construct, the tetramerization domain was replaced with a \emph{de novo} two-helix hairpin that was designed to assemble into a water-soluble octameric pore (WSHC8 from \cite{xu_computational_2020}, PDB 6O35) and the Halotag is N-terminal to the motor. Constructs were cloned into the insect expression vector pBiEx-1.

For protein expression, plasmids were directly transfected into Sf9 cells as described previously \cite{liao_engineered_2009}. Purification was performed as described in \cite{liao_engineered_2009} and \cite{ruijgrok_optical_2021}. Briefly, proteins were purified using anti-Flag resin and labeled with Alexa Fluor 660 HaloTag Ligand (Promega). Proteins were eluted into storage buffer containing glycerol and then immediately flash-frozen in small aliquots and stored at --80$^\circ$C until use.

\subsubsection{Assay Conditions}
Actin filaments were polymerized at a 1:10 labelling ratio and a concentration of 2 \(\mu\)M in a 50 $\mu$L polymerization mix. This mix contained 1X F-buffer [10 mM imidazole, 1 mM MgCl2, 50 mM KCl, 0.2 mM egtazic acid (EGTA), pH 7.5] with each of the concentrations of ATP studied. No additional MgCl2 was added with ATP. To minimize photobleaching, an oxygen scavenging system (4.5 mg/mL glucose, 2.7 mg/mL glucose oxidase (catalog no.\ 345486, Calbiochem, Billerica, MA), 17000 units/mL catalase (catalog no.\ 02071, Sigma, St. Louis, MO) and 0.5 vol. $\%$ $\beta$-mercaptaethanol was added. Actin filaments were crowded to the surface by including 0.3\% w\% 400 cP methylcellulose in the polymerization mix. Capping protein was first thawed on ice, then diluted to 500 nM in 1X F-buffer, and added at a final concentration of 30 nM in the mix. This polymerization reaction was allowed to proceed for one hour on ice before it was added to the imaging chamber. 

The imaging chamber was created by first rinsing a small glass cloning cylinder (catalog no.\ 09-552-20, Corning Inc.)\ with ethanol and then attaching it to a silanated glass coverslip with two-part epoxy. To prevent the actin from sticking and maintain fluidity, the coverslip was coated with a thin layer of Novec 7500 Engineered Fluid (3M, St. Paul, MN) that included PFPE-PEG-PFPE surfactant (catalog no.\ 008, RAN Biotechnologies, Beverly, MA) at 2\% w/v before the polymerization mix is added. The mixture was allowed to sit in the sample chamber for about 30 min before imaging to allow for the formation of the nematic.

The sample was imaged on an Eclipse-Ti inverted microscope (Nikon, Melville, NY) in confocal mode utilizing a spinning disk (CSU-X, Yokagawa Electric, Musashino, Tokyo, Japan) and a CMOS camera (Zyla-4.2 USB 3; Andor, Belfast, UK). Experiments were imaged at one frame every 2 s.

\subsection{Data analysis}

Flow fields were calculated between every two frames from time lapse images with optical flow using the Classic+NL-fast method \cite{sun_secrets_2010,sun_quantitative_2014}. This method is based on the classic Horn–Schunck method which minimizes an objective function penalizing intensity differences between subsequent frames (the data term) as well as enforcing smoothness in the estimated field. Flow is estimated at various spatial scales iteratively to capture first global and then local motion. The optical flow code was obtained from https:\slash \slash cs.brown.edu\slash people\slash mjblack\slash code.html.

Average flow speed \(v\) was calculated from the \(N\) vectors, \(u_i\), as \(v = \sum |u_i|/N\). The velocity correlation length quoted in Figure S2 was calculated as the distance \(r\) at which the velocity autocorrelation function \(C_{uu}(r) = \langle u_i(0) \cdot u_j(r)/|u_i| |u_j|\rangle \) reaches \(1/e\), where the average is over all pairs $(i,j)$ and \(e\) is Euler's number.

\(\ell_\text{vort}\) was calculated with the method of correlated displacement fields, as described in \cite{molaei_measuring_2023}. Briefly, the normalized cross correlation is measured in two dimensions between the vorticity field \(\nu\) and the velocity field \(u\). This procedure effectively measures the response of the nematic to a unit vortical perturbation at the origin. To extract a length scale from this response, the azimuthal average of the correlation field is taken. This average results in a one dimensional function with a single maximal extreme. \(\ell_\text{vort}\) is the distance \(r\) at which this maximum occurs. This length scale has been shown in active nematics to be equal to the average radius of a vortex in the flow field \cite{molaei_measuring_2023}. Error for this method was calculated by measuring \(\ell_\text{vort}\) over 5 separate non-overlapping sets of frames from the 100 s of steady-state data considered in \(v_\text{rms}\). The code is available at https://github.com/Gardel-lab/ResponseFunction.

\subsection{Motor Stepping Model}

The code to run an analyze the myosin stepping model described in Results is available at \\https://github.com/Gardel-lab/myosin\_stepping\_model.

\subsection{Lattice Boltzmann Simulations}

Simulations of active nematic hydrodynamics were performed using a custom Julia implementation of the hybrid lattice Boltzmann algorithm \cite{floyd_simulating_2023, carenza_lattice_2019}.  The simulated equations of motion are the same as those detailed in \cite{zhang_spatiotemporal_2021, colen_machine_2021}.  The simulation domain consists of 400$\times$400 lattice points in two dimensions with periodic boundary conditions.  The turbulent state was generated by initially perturbing the system and evolving for $15,000$ steps, and then data was collected every $50$ steps for another $15,000$ steps.  For each condition we ran 5 independent trials using different random seeds for the initial perturbation.  We used the following parameters (in lattice units): a collision time $\tau= 1.5$ (corresponding to viscosity $\eta = 1/3$), a flow-alignment parameter $\xi = 0.7$, a rotational diffusion constant $\Gamma = 0.13$, and polarization free energy coefficients of $A_0 = 0.1$, $U = 3.5$, leading to an equilibrium nematic polarization magnitude of $q = 0.62$.  The elastic constant $K \in [0,0.1]$ and activity coefficient $\alpha \in [0,0.01]$ (where positive $\alpha$ corresponds to extensile activity) were varied to generate the results shown here.

\bibliographystyle{unsrt.bst} 
\textnormal{\bibliography{references.bib}}

\newcommand{\SItext}{%
  \section*{Supporting Information Text}
  \stepcounter{SItext}
}
\renewcommand{\thefigure}{S\arabic{figure}}
\renewcommand{\theequation}{S\arabic{equation}}
\setcounter{figure}{0} 
\setcounter{equation}{0}
\clearpage
\section{Supplemental Figures}

\begin{figure}[h]
    \centering
    \includegraphics[width=0.8\textwidth]{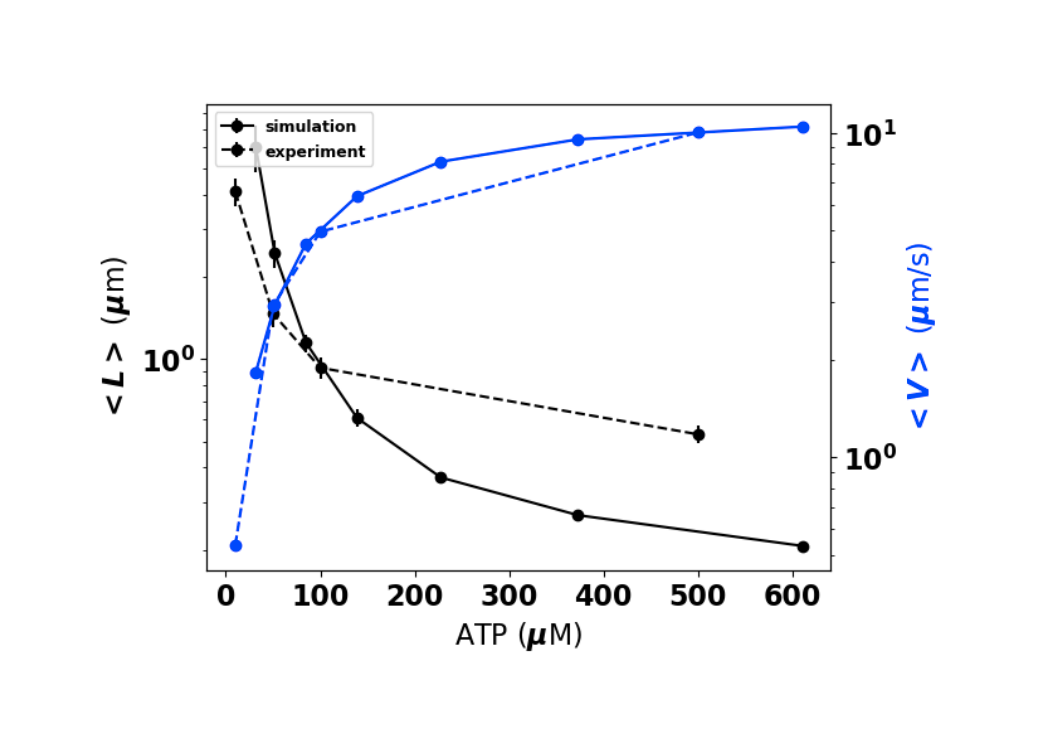}
    \caption{\textbf{Simulations reproduce single filament velocity and run length trends.} Single filament motor velocity (blue) and single filament run length (black) from experiments (dashed lines) and simulaitons (solid lines) over a range of ATP concentrations. The final rates we compute after 10,000 tuning steps are 1821 s$^{-1}$, 932 s$^{-1}$, 6 s$^{-1} \mu$M ATP.}
    \label{model_controls}
\end{figure}

\begin{figure}[h]
    \centering
    \includegraphics[width=0.8\textwidth]{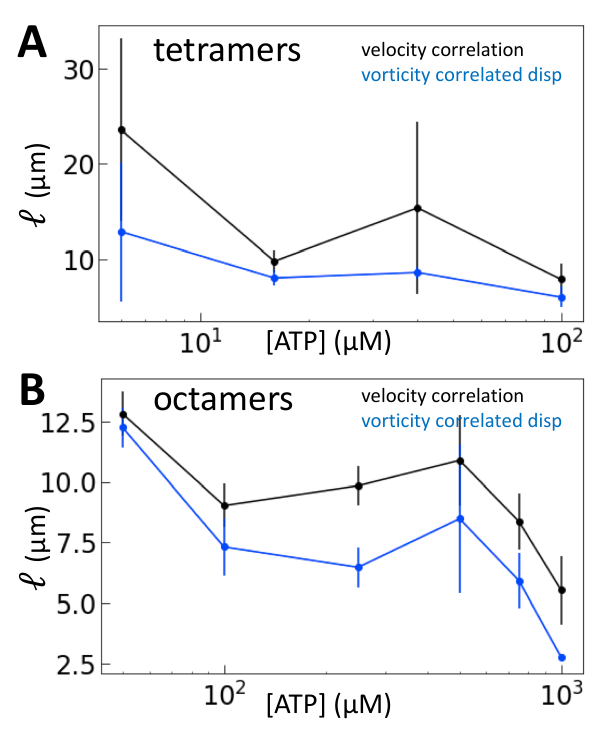}
    \caption{\textbf{$\ell_{vort}$ robustly captures nematic length scale.} Comparison of $\ell_\text{vort}$ as calculated in \cite{molaei_measuring_2023} and the traditional velocity correlation length --- $C_{vv} = 1/e$ --- for nematics driven by 120 pM tetramers (A) and 50 pM octamers (B). Errorbars are averages over five separate frames for $C_{vv}$ and five 10s snippets for $\ell_{vort}$.}
    \label{length_scale_controls}
\end{figure}

\begin{figure}[h]
    \centering
    \includegraphics[width=0.8\textwidth]{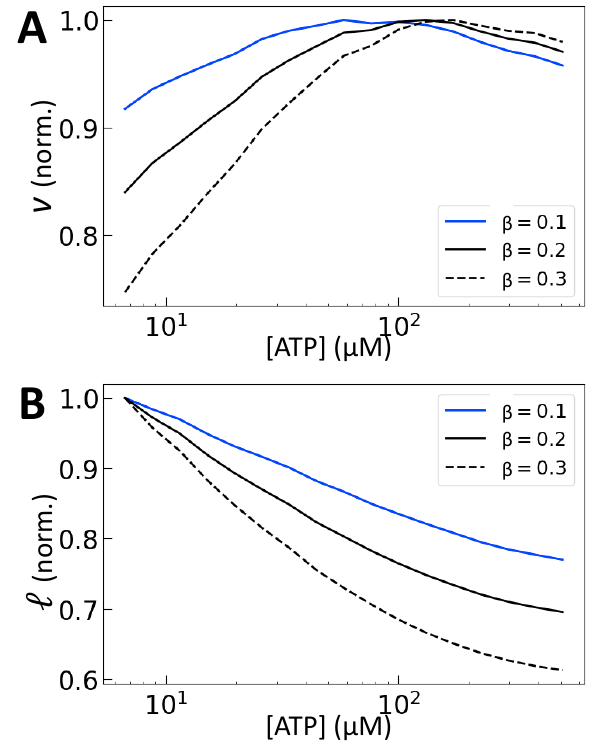}
    \caption{\textbf{Strong activity coupling decreases nonmonotonicity in tetrameric driven nematics.} \(v\) (A) and \(\ell\) (B) from scaling predictions for different coupling exponents \(\beta\).}
    \label{beta}
\end{figure}

\begin{figure}[h]
    \centering
    \includegraphics[width=0.8\textwidth]{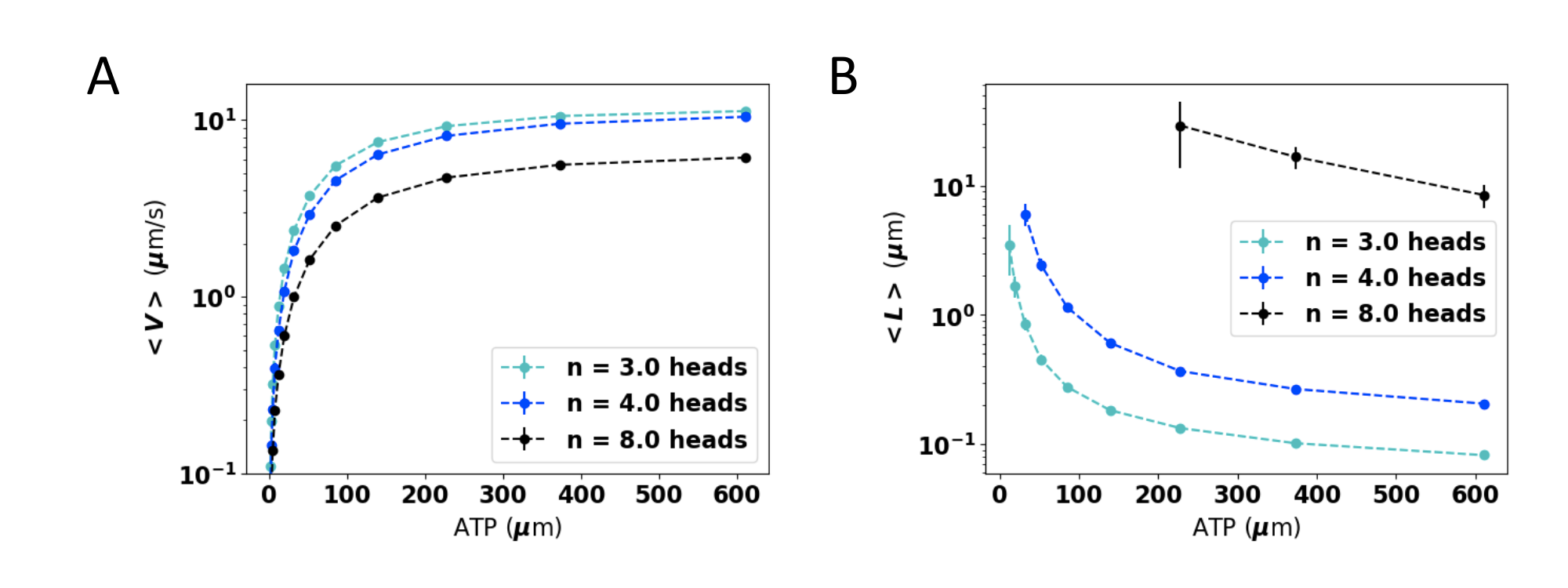}
    \caption{\textbf{Motor velocity and run length are cluster valency dependent phenomena.} Single filament motor velocity (A) and single filament run length (B) measured from simulation for clusters with 3,4, or 8 heads per cluster.}
    \label{cluster_valency}
\end{figure}

\begin{figure}[h]
    \centering
    \includegraphics[width=0.8\textwidth]{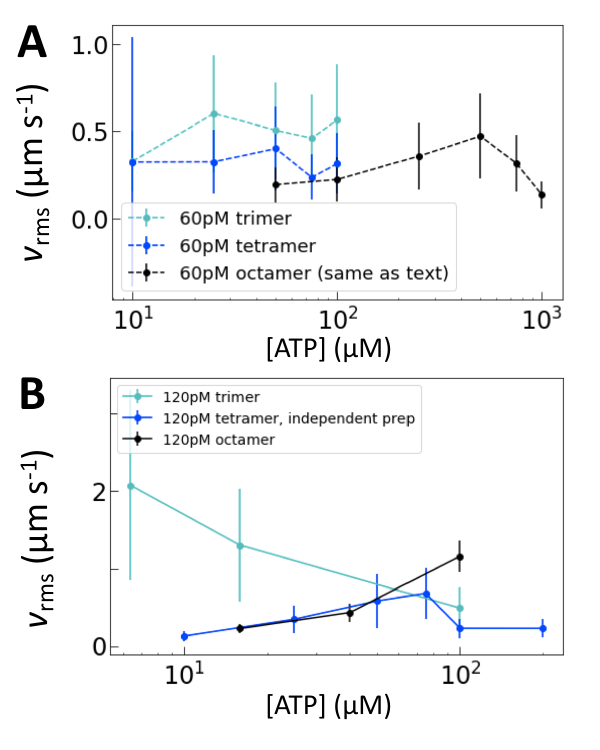}
    \caption{\textbf{Peak shift is robust across days and motor concentrations.} $v_\text{rms}$ for independent replicates of oligomerization data at $60$ pM (A) or $120$ pM motor clusters. All data are different from those in the text except for the octamers in (A) which are included as a reference. Error bars are standard deviations of $v_\text{rms}$ over 100 s of steady-state activity.}
    \label{replicates}
\end{figure}

\begin{figure}[h]
    \centering
    \includegraphics[width=0.8\textwidth]{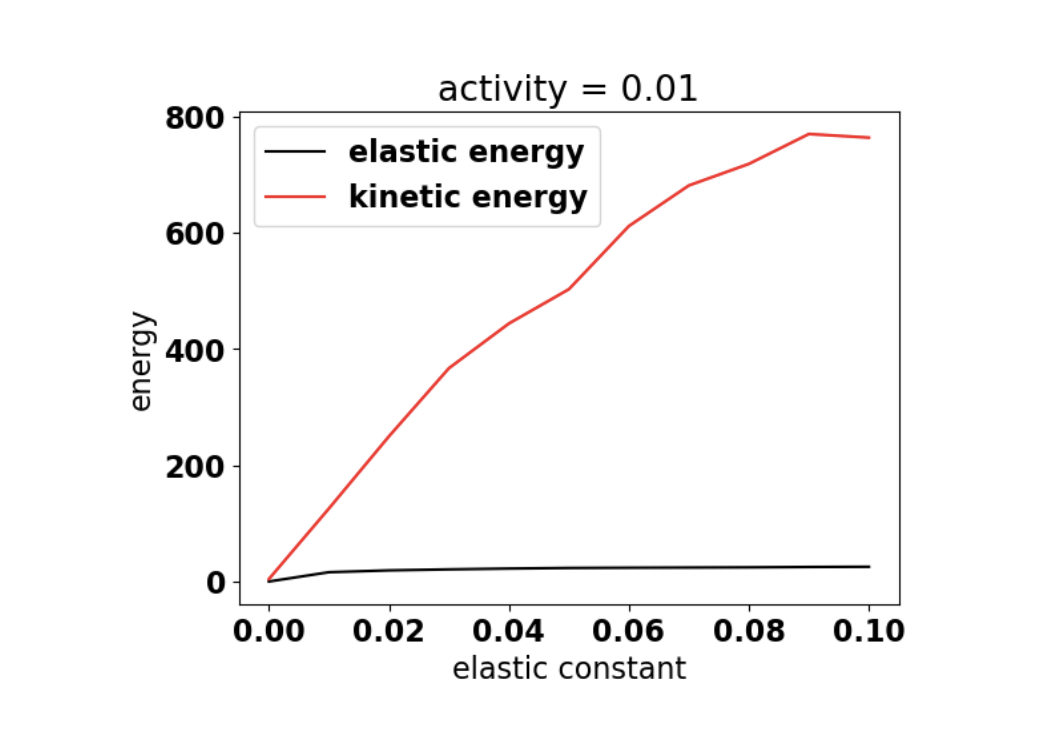}
    \caption{\textbf{Nematic elasticity increases energy in the nematic.} Elastic (black) and kinetic (red) energy for nematics in lattice Boltzmann simulations with constant \(\alpha\) = 0.01 across a range of $K$.}
    \label{LBE}
\end{figure}

\end{onehalfspacing}
\end{document}